
\magnification= \magstep 1
\hsize = 17 truecm   \vsize = 23 truecm
\def\singlespace{\baselineskip=12pt}
\def\doublespace{\baselineskip=14pt}
\def\sesquispace{\baselineskip=18pt}
\def\nl{\hfill\break }         
 at 10truept 
 at 10truept 
\def\startindent{\leftskip=1.5truecm\rightskip=1.5truecm}
\def\endindent{\leftskip=0truecm\rightskip=0truecm}
\newskip\footskip\footskip14pt plus 1pt minus 1pt 
\def\footnotefont{\it}\def\f@t#1{\footnotefont #1\@foot}
\def\f@@t{\baselineskip\footskip\bgroup\footnotefont\aftergroup\@foot\let\next}
\setbox\strutbox=\hbox{\vrule height9.5pt depth4.5pt width0pt}
\global\newcount\ftno \global\ftno=0
\def\foot{\global\advance\ftno by1\footnote{$^{\the\ftno}$}}
\global\newcount\refno
\def\ref#1{\xdef#1{[\the\refno]}
\global\advance\refno by1#1}
\global\refno = 1
\global\newcount\meqno
\def\eqn#1#2{\xdef#1{(\secsym\the\meqno)}
\global\advance\meqno by1$$#2\eqno#1$$}
\xdef\secsym{}\global\meqno = 1
\def\half{{\textstyle{1\over 2}}}  
\def\quart{{\textstyle{1\over 4}}} 
\def\Tr{\hbox{Tr}}           
\def\to{\rightarrow}         
\def\tilde{\widetilde}       
\def\bbar{\overline}         
\def\Complexes{{\rm C}\llap{\vrule height6.3pt width1pt depth-.4pt\phantom t}}
\def\epsi{\epsilon\!\!\!_{_\sim}\,}
\def\V{\ V\!\!\!\!\!_{_\sim}\  }
\def\E{\tilde E}
\def\NP#1{Nucl. Phys. {\bf #1}}

\def\PR#1{Phys. Rev. {\bf #1}}
\def\PRL#1{Phys. Rev. Lett. {\bf #1}}
\def\NC#1{Il Nuovo Cim. {\bf #1}}
\def\CQG#1{Class. Quantum Grav. {\bf #1}}
\input epsf
\sesquispace
\rightline {DFUP 112-95, gr-qc/9512040}\par
\bigskip\bigskip
\centerline{\bf Quantizing Regge Calculus.}
\bigskip
\singlespace
\centerline {  Giorgio Immirzi}
\bigskip
\noindent Dipartimento di Fisica, Universit\'a di Perugia and
Istituto Nazionale di Fisica Nucleare, sezione di Perugia,
 via A. Pascoli, 06100 Perugia, Italy. E--mail: immirzi@perugia.infn.it
\bigskip
\singlespace
\startindent
\centerline{\bf Abstract}
\medskip
\noindent A discretized version of canonical gravity in $(3+1)$--d introduced
in a previous
paper is further developed, introducing the
Liouville form and the Poisson brackets, and studying them in detail in an
explicit
parametrization that shows the nature of the variables when the second class
constraints are imposed. It is then shown that, even leaving aside the
difficult question of imposing the first class constraints on the states,
 it is impossible to quantize the
model directly, using complex variables and  leaving  the second class
constraints to fix the metric of the quantum Hilbert, because one cannot find
a metric which makes the area variables hermitean.
\bigskip
\endindent
\doublespace \bigskip \medskip
\noindent{\bf 1. Introduction. }\medskip\nobreak\medskip
In a previous paper\ref\gitwo\  I discussed a possible way
of discretizing canonical gravity in 3+1 dimensions, dividing space
in tetrahedra, and associating pairs of variables to the
triangles that separate them:  the oriented area of the triangle,
and the Lorentz transformation that takes us from one tetrahedron
to the next through the triangle.
 This scheme is not very different from others recently proposed
\ref{\leerent}\ref{\renate}(these authors use a cubical lattice,
but that perhaps is not very important), but draws rather from
the well known work of T. Regge\ref\regge\ to adapt to a discretized
theory the new variables of A. Ashtekar\ref\abhay .
If one could quantize this particular form
of "Regge calculus" one would have a  quantum gravity which is
regularized,  yet is faithful to the geometrical spirit of
Einstein's theory; in particular, unlike other regularizations,
does not require an arbitrary background metric.

With this aim in mind, this program is continued in this paper
specifying and analysing in detail the Liouville form, and
therefore the Poisson  brackets between the basic variables.

{}From this one should go
on to eliminate second class constraints (perhaps
introducing Dirac brackets),  check the algebra between
first class constraints, finally proceeding to the quantization
of the theory. However, it is central to
the quantization program based on the Ashtekar variables the
idea that one does not solve the second class constraints (which
are reality conditions on the complex basic variables), but uses
them to fix the metric in the Hilbert space of the quantized
theory.  In this spirit I have ignored the difficult problem of
checking whether the algebra of the constraints closes, but rather
attempted  a straight conversion of Poisson brackets into
commutators, and checked whether one can find an acceptable scalar
product in the Hilbert space. The result is that no, there cannot
be a scalar product which makes the basic area variables
hermitian.  Thus the procedure which was shown to work fine for linearized
gravity \ref\ars\  fails in this case, and so does the attempt to
produce a discretized version of the  "net states" of C. Rovelli
and L. Smolin \ref{\carlolee} or of their volume and area
operators. I am not able to say whether this should be taken as
evidence against the viability of their program, or of the
quantization program in general, but find little reason for
optimism. And, in a way, one should not be too surprised that a
relatively na\"\i ve approach like this fails, considering how tricky
is the construction of the scalar product in the simpler case of
$2+1$--d gravity\ref{\abhayrenate}.

The alternative, real $SU(2)$ connection
introduced by J. F. Barbero\ref{\nando} is not a Levi--Civita connection, and
therefore does not have an obvious geometric interpretation in this discretized
context. Nor can one hope to start with an $SU(2)$ connection and recover the
full theory by analytic continuation, unless one relaxes the constraints that
come from the geometric interpretation. It is plausible to conjecture that
the imaginary part of the Liouville form vanishes when the Gauss law and the
reality conditions are imposed,  and therefore if the group were restricted to
$SU(2)$ there would be no local degrees of freedom, and nothing to
analytically continue.

I have tried (but failed) to prove this conjecture using an
explicit real parametrization of the basic canonical variables that makes
clear the meaning of the reality conditions; numerically,
 on a simple
5--tetrahedra model, the conjecture appears to be true.
The Liouville form, if its imaginary part is indeed zero
depends only on the areas of the triangles and the rapidities of
the Lorentz transformations, very much like the one proposed by
t'Hooft\ref\thooft\  for his discretized 2+1 dimensional theory.

The discretization scheme is briefly summarized in \S 2, while
in \S 3 the Liouville form and the Poisson brackets are introduced.
The Liouville form is further discussed in \S 4, using an explicit, real
parametrization, and the restriction
imposed by the reality conditions are analysed in detail.
Finally in
\S 5  I discuss the quantization of the theory and  the reasons
why it is not possible to construct a scalar product that make the
area variables hermitean.

\bigskip\noindent{\bf 2. Outline of the model. }\medskip\nobreak\medskip
Following the basic ideas of Regge calculus, 3--space is
divided  in tetrahedra, each with a flat inside. We may therefore
choose an inertial frame for each tetrahedron, and expect that a
Lorentz transformation will be needed to go from one tetrahedron
to the next, and that because of curvature one will end up in a
different Lorentz frame going round an edge.

In terms of Ashtekar variables, this intuitive picture
means that we have in each tetrahedron some constant
$\E^{ia}$, and a complex connection $A^i_a$ with support on the
triangles separating tetrahedra.
So, if tetrahedron A ((1234) in fig.1) shares the triangle (123)
with its neighbour B (1235), we can associate to the triangle
variables $(S,g)$ given in local coordinates by:
\eqn\uno{S_A:= \tau_i S^i_A := \tau_i \E^{ia}_A\epsi_{abc}S^{bc}=
 -g(AB)\,S_B\,g(AB)^{-1} \ ;\qquad  g(AB) :=P\exp\int_A^B A\cdot\tau\, dl }
where $\tau_i={\sigma_i\over2i}$, $g(AB)$ is the
$SL(2,\Complexes )$ element that takes from the B to the A frame
across the triangle, $S_A^i$ (or the traceless $2 \times 2$ matrix
$S_A$) is the (oriented) area of the triangle in the frame of A;
if the vertices of the triangle are $(x_1^a,x_2^a,x_3^a)$,
$S^{ab}=\half (x_1x_2+x_2x_3+x_3x_1)^{[ab]}$.
The areas are oriented outwards in each frame,  which explains
the minus sign in \uno . Curvature is found going round an edge, e.g.:
\eqn\iii{R_{(12)A}:=g(AB)g(BC)g(CA):=\exp (F_{(12)A})   }
\epsfxsize=300pt
\epsfbox{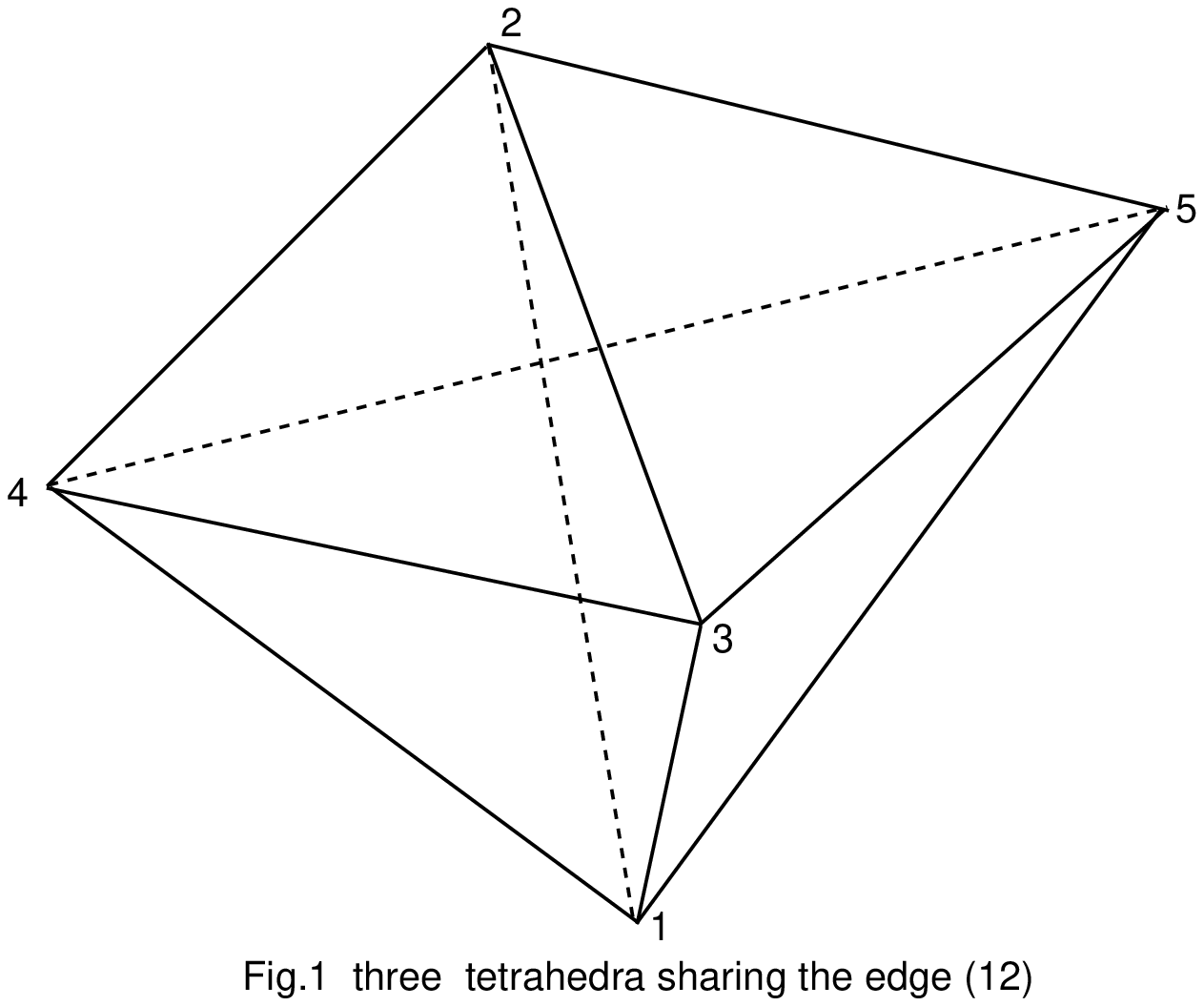}

Gauge transformations are Lorentz ($SL(2,\Complexes )$)
transformations in each tetrahedron:
\eqn\ii{S_A\to g_AS_Ag_A^{-1}\ ,\ \  S_B\to g_BS_Ag_B^{-1}\ ,\ \
 g(AB)\to g_Ag(AB)g_B^{-1} }
{}From this equations we see that the 3-vectors $S^i$ trasform
according to the self-dual (or $(1,0)$) representation of the
Lorentz group. It is the use of this representation that assures the nice match
between the formalism and the geometric picture. The price to pay
is that the variables are complex, subject to
"reality conditions" mutuated from the continuum theory \abhay .
The $S^i$ variables are in general complex, but
a first set of conditions is that in a tetrahedron all $S^2$  must
be real  positive and all scalar products real; this  insures that
we can choose a frame such that all the $S^i$ are real vectors
("time gauge"). To express the second set of conditions, which involve the
$g$ variables,
we have to define for each pair of triangles belonging to
a tetrahedron (say (123) and (214) of A in fig.1) a
variable $\tilde l$, associated to their common edge  by:
\eqn\iv{  [S_{(123)A},S_{(214)A}] ={\textstyle 3\over 4}\V_A
[\tau_i,\tau_j]\epsi_{abc}\tilde E^{ia}_A\tilde E^{jb}_A (x_2-x_1)^c
  := {\textstyle 3\over 2} \V_A \tilde l_{(12)A} }
where $\V_A = {\textstyle 1\over
6}\epsi_{abc}(x_1x_2x_3+x_2x_1x_4+x_3x_4x_1+x_4x_3x_2)^{abc}$
is the volume of A in local coordinates.
In the same way the triangles $(123),\;(215)$ give a variable
$\tilde l_{(12)B}$. If $g(AB)$ were just $1$, the three vectors
$S_{(123)},S_{(214)},S_{(215)}$ would be coplanar; this basic
fact is generalized to:
\eqn\v{[\tilde l_{(12)A},g(AB)\,\tilde l_{(12)B}\,g(AB)^{-1}]=
 i\, c\, S_{(123)A}       }
for some real constant $c$, and this is the second set of reality
conditions.

Apart from these conditions, which are second class constraints, we
have first class constraints, again mutuated from the continuum
theory. First of all, the tetrahedra must close, so, for
tetrahedron A with $I =(123),(231),(341),(432)$:
\eqn\vi{\sum_I S_{I\, A}=0}
which is our "Gauss law". Further, at a more tentative level, we
have analogues of the vector and the scalar constraints; for each
tetrahedron:
\eqn\vii{ \sum_l \epsilon_{acb}F^c_l \tilde l_l^b
   =0\ ;\qquad \sum_l F^a_l \tilde l_l^a =0}
these expressions were shown to give the Ashtekar constraints in
the continuum limit, and to be respectively pure imaginary  and
real using the reality conditions \v .

This is the setting on which I propose to build a quantum theory.

\bigskip\noindent{\bf 3. Poisson brackets. }
\medskip\nobreak\medskip
To put some dynamics in the kinematic setting, we need first of
all to specify a Liouville form, so that the Poisson brackets
between the variables can be calculated.

The form I propose is, labeling triangles and tetrahedra with
indices $I$ and $T$ respectively:
\eqn\viii{\Theta = i\sum_T\sum_{I\in T} \Tr (S_{I\,T}\, dg_I\,g_I^{-1}) }
This is similar to what other authors have proposed [2,3], and
looks very much like the Liouville form for a set of rotating tops
(but notice the factor $i$, and that the $g_I$ are $SL(2,\Complexes
)$ matrices, not $SU(2)$). One can show that \viii reproduces the
continuum Liouville form in a na\"\i ve continuum limit. In fact,
a given triangle contributes twice to the sum, but the two
contributions are equal; explicitely for say (123) of fig.1:
\eqn\ix{i\Tr (S_A\, dg(AB)\,g(AB)^{-1})+i\Tr (S_B \,dg(BA)\,g(BA)^{-1})=
 2i\Tr (S_A\, dg(AB)\,g(AB)^{-1})}
We may split the displacement from the middle of A to the
middle of B in two parts, so that
\eqn\x{g(AB)\approx 1 + \tau_i A^i_a (\delta^a_A+\delta^a_B)}
from this approximate form and \uno\ we get for the contribution of
a tetrahedron to $\Theta$:
\eqn\xi{\Theta_T=i\sum_{t\in T}\Tr (\tau_i \E^{ia}\epsi_{abc}S_t^{bc}\tau_j
dA^j_d\delta_{tT}^d )= -i\half\E^{ia}dA^i_d \epsi_{abc}\sum_{t\in T}S_t^{bc}
\delta_{tT}^d = -i\E^{ia}dA^i_d \V_T }
which agrees with the continuum theory \abhay ; in the last
step we have used an identity (eq.31 of \gitwo ) for the volume
$\V_T$ of $T$.

The other merit of the proposed form is that under gauge trasformations:
\eqn\xii{\Tr (S_A\, dg(AB)\,g(AB)^{-1})
 \to  \Tr (S_A\, dg(AB)\,g(AB)^{-1}) +\Tr (S_A\,dg_A\, g_A^{-1})
+\Tr (S_B\,dg_B\, g_B^{-1})}
summing over all triangles and all tetrahedra we see that if the
Gauss law \vi\ is satisfied, the Liouville form is invariant, just
like in the continuum.

The sympletic form is $\Omega =d\Theta$, and the algebra to
calculate the Poisson brackets can be borrowed from the theory
of the rotating top. The result is (I use the notation
$g^{(1)}=g\otimes  1,\ g^{(2)}=1\otimes g$):
\eqn\xiii{\{ S_A^i,S_A^j\}=-i\,\epsilon_{ijk}S_A^k\ ;\quad
\{g(AB), S^i_A\} =-i\,\tau_i\cdot g(AB)\ ;\quad
\{ g(AB)^{(1)},g(AB)^{(2)} \} =0}
We see that the $S$ variables (summed for each tetrahedron) are the
generators of gauge transformations. The factor $i$ may appear
surprising, and it will prove fatal to the na\"\i ve quantization scheme
of \S 5,
 but our insisting that the Liouville form has the
correct continuum limit implies that it must be inexorably there.
In the continuum, when the Gauss law and the reality conditions
are satisfied and in time gauge, the Ashtekar connection becomes
the Sen connection:
\eqn\xiv{A^i_a=-\half\epsilon_{ijk}\Omega^{jk}_a+ie^{ib}K_{ab}}
from this we see that the  $i$ that appears in the last member of
\xi\ is indeed correct, because the dynamics is in the imaginary part of the
Ashtekar connection, the extrinsic curvature $K_{ab}$
($\Omega^{jk}_a$ is the Levi--Civita connection).

\filbreak
\bigskip
\noindent{\bf 4. Geometrodynamics.  }\bigskip
Given that the presentation of the variables and of the model has
been rather formal, one may wonder whether the details
are so important: could one change the Poisson brackets removing
the offending factor $i$? or solve the reality conditions and use
real variables, like in geometrodynamics? Removing the factor $i$
would mean to start with a model in which the group is $SU(2)$,
from which the full theory could be reached by analytic
continuation. I think this unlikely: as we shall see, taking an
explicit real parametrization for the canonical variables, one
finds that it is unlikely that a geometrically sensible
$SU(2)$ theory exists, because its Liouville form would most
likely  vanish, and there would be no local degrees of freedom left.

Choosing the time gauge for tetrahedra A and B, i.e. frames in
which all the $S^i$ are real 3--vectors, we can introduce the
following parametrization for the basic variables:
 \eqn\xxiv{S_A :=u_A\cdot\tau_3s\cdot u_A^{-1}\ ;\quad
   S_B:=u_B\cdot\tau_3s\cdot u_B^{-1}\ ;\qquad {\rm with:}\ \
u_A=e^{\alpha \tau_3}e^{\beta \tau_2}\ ;\quad
u_B:= e^{\gamma \tau_3}e^{\delta \tau_2}}
where $\alpha\ldots\delta$ are the polar angles of $S^i_A,\; S^i_B$
and $s>0$. Then by eq.\uno\ $g(AB)$ must be of the form:
\eqn\xxv{ g(AB):=u_A e^{(\Phi +i\zeta )\tau_3}2\tau_2u_B^{-1}  }
Here $\zeta$ is the rapidity of the Lorentz transformation; the
matrix $2\tau_2$ is there because the $S$ have opposite orientation
in the two frames, and $\Phi$ determines the axis around which the
rotation by $\pi$ takes place. Notice that this is essentially the
parametrization used in \ref\fad .

Expressed in terms of these variables the contribution of a
triangle to the the Liouville form becomes:
 \eqn\xxvi{\eqalign{ 2i\Tr (S_A\,dg(AB)\,g(AB)^{-1})&= s\,d\zeta
-i\,s\,d\Phi + 2is\,\Tr (\tau_3u_A^{-1}du_A)  +2is\,\Tr
(\tau_3u_B^{-1}du_B)=\cr &=s\,d\zeta -i\,s\,d\Phi-i\,s\,\cos\beta
d\alpha -i\,s\,\cos\delta  d\gamma \cr}}
The real part of this form is just what was assumed in \thooft\
in the $(2+1)$--d case, which is promising. As for the imaginary
part: in the continuum, when the Gauss law and reality condition
are imposed,  the imaginary part of the Liouville form involves the
Levi--Civita connection (see eq.\xiv ), and it integrates to zero.
I therefore conjecture that, under the same hypothesis,
\eqn\xxvibis{Im\,\Theta =\sum_{t}s_t(d\Phi_t+\cos\beta_t d\alpha_t
+\cos\delta_t d\gamma_t ) =0 \ \ ?}
This is a crucial test of consistency for the whole scheme,
but I have not been able to prove that
indeed it is so. It is however very plausible that
when all the $\zeta$ are set to zero we are left with the
Levi--Civita connection. And indeed,
replacing eqs.\xxiv\xxv\ in the reality conditions \v ,
one can determine the angle $\Phi$ in
terms of the others. To keeps things simple,
define, with reference to the edge $(12)$ of fig.1:
\eqn\xxvii{\eqalign{u^{-1}_{(214)A}u_{(123)A}&:=\exp
(\tau_3\phi_{(12)A})\exp (\tau_2\theta_{(12)A})
\exp (-\tau_3\psi_{(12)A})\cr
u^{-1}_{(213)B}u_{(125)B}&:=\exp (\tau_3\phi_{(12)B})
\exp (\tau_2\theta_{(12)B})\exp (-\tau_3\psi_{(12)B})\cr}}
where  $\theta_{(12)A},\ \theta_{(12)B}$ are the angles between
the faces $(123)$ and $(214)$, and between $(213)$
and $(125)$. After some algebra, we find that eq.\v\ for the
edge $(12)$  and  tetrahedra A and B becomes:
\eqn\xxviii{ \cosh\zeta \sin\theta_{(12)A} \sin\theta_{(12)B}
\sin (\phi_{(12)B}+\psi_{(12)A}-\Phi_{(123)}) + i\,\sinh\zeta
(\ldots ) =  i\,c }
for some real c, so that:
\eqn\xxix{\Phi_{(123)}=\phi_{(12)B}+\psi_{(12)A}\ \ {\rm mod}\ \pi}
This simple relation implies that the angles $\Phi$ neatly
cancel from the expression for the curvature, which becomes:
\eqn\xxx{\eqalign{
 R_A &= u_{3A}e^{(\Phi_3+i\zeta_3)\tau_3}\,2\tau_2\, u^{-1}_{3B}
u_{5B}e^{(\Phi_5+i\zeta_5)\tau_3}\,2\tau_2\,u^{-1}_{5C}
u_{4C}e^{(\Phi_4+i\zeta_4)\tau_3}\,2\tau_2\,u^{-1}_{4A}=\cr
&=u_{4A}e^{\phi_A\tau_3}e^{\theta_A\tau_2}e^{i\zeta_3\tau_3}\,
2\tau_2\, e^{\theta_B\tau_2}e^{i\zeta_5\tau_3}\,2\tau_2\,
e^{\theta_C\tau_2}e^{i\zeta_4\tau_3}\,2\tau_2\, e^{-\phi_A\tau_3}
u_{4A}^{-1} \cr}  }
(we have dropped subscripts $(12)$), in agreement with the
expression found in \gitwo .

The reality conditions eq.\v\ also give more relations
between the angles, because  we should obtain the same
$\Phi_{(123)}$ repeating  the argument that leads to eq.\xxix\
for the edges $(23)$ and $(31)$; therefore:
\eqn\xxxi{\phi_{(12)B}+\psi_{(12)A}=\phi_{(23)B}+\psi_{(23)A}
=\phi_{(31)B}+\psi_{(31)A} \ \ {\rm mod}\ \pi}
These relations cannot be all independent, although I have found
the detailed trigonomety too complicted to disentangle. One can
however argue as follows:  suppose the Regge lattice has $N_0$
vertices, $N_1$ edges, $N_2$ triangles and
$N_3$ tetrahedra. If it is the discretization of a closed
3-manifold we know that \ref\frolich :
\eqn\xxxii{2N_3=N_2\quad ;\quad N_0-N_1+N_2-N_3=0}
The metric and the orientation of each frame requires that we give
the lenghts of the $N_1$ edges and three angle for each of the
$N_3$ tetrahedra. Alternatively, we can give the $N_2$ areas and
$8\,N_3$ angles, subject to the $3\,N_3$ condition that each
tetrahedron closes (eq.\vi ) and to $X$ conditions coming from the
reality conditions eq.\xxxi . Therefore:
\eqn\xxxiii{N_1+3N_3=N_2+8N_3-3N_3-X \quad\Longrightarrow\quad
X=2N_2-N_1 }
Hence, between the $2N_2$ relations eq.\xxxi\ we see that
only $2N_2-N_1$ are independent. This is plausible if we remember
that these relations basically state the coplanarity between the
triangles that share an edge. But just because of this geometric
motivation, it is difficult to imagine a theory in which the
$\Phi$ angles  would be independent variables, and the angular
contribution to the Liouville form, i.e. its
imaginary part, is most likely to vanish as conjectured above,
eq.\xxvibis .

Having failed to prove eq.\xxvibis , I have tried numerically
to check whether it is true at least in the
simplest model: $S^3$ divided in five tetrahdra, i.e. the boundary of a
4--simplex
\ref\ruth . One may assign random lengths to the 10 edges, choose arbitrarily
a frame in each tetrahedron and calculate all the areas and angles; then vary
slightly the lengths, recalculate everything and in this way estimate
$Im\,\Theta$ when the Gauss
law and the reality conditions are imposed.
However carefully one repeats the calculation, the extent to which it can be
trusted
is debatable, but it does confirm the conjecture;
more precisely one finds that eq.\xxvibis\ holds, numerically, more or less
like
the Regge  theorem (see \regge , app.1) that the contribution of the angle
variation to the variation of his action is zero,
$\sum l_i d\Theta_i=0$. However, while the Regge theorem holds for each
tetrahedron, it does not appear that, if one  uses
eq.s\xxix\xxxi\  to reorganize the sum in eq.\xxvibis\ as a sum over tetahedra,
the indidual contributions vanish.

I would like to conclude this section with one further question: when all
constraints
are taken into account, how many degrees of freedom does the theory have?
This is an important, and non trivial question, first put by Regge in
\regge , but I do not know of an articulate answer. We would like to know how
many of
the $4N_3$ constraints eq.\vii\ are independent, and by how much do they reduce
the
number of independent variables. From areas and angles,
taking the relevant constraints into account, we get $N_1$ independent
variables , to which one should add the $N_2$ rapidities
$\zeta$. As a first guess, I would expect $N_1$ relations between these to come
from  the  constraints eq.\vii , leaving us with $N_2=2N_3$ independent
variables,
simply because I see no other way of getting an even number. The analysis of
the
constraints is certainly not going to be easy.
\bigskip\noindent{\bf 5. Quantization. }\medskip\nobreak\medskip
Given the Poisson brackets, one should  go on to check the
algebra of the constraint, a very difficult task, that I have
not attempted. On the other hand, one may simply hope for the best,
and try to quantize the theory by blindly turning Poisson
brackets into commutators
 \eqn\xv{[ \hat S^i,\hat S^j ]
=-\epsilon_{ijk}\hat S^k\ ;\quad [\hat S^i,\hat g]=\tau_i\cdot
\hat g\ ;\quad [\hat g^{(1)},\hat g^{(2)}] =0
}
The form of these commutation relations makes the connection
representation appear "natural":  we take wave functions to be
holomorphic, gauge invariant functions of the $g$--s. On this wave
functions the $S^i$-s act like (minus) the generators of the
left--regular representation of $SL(2,\Complexes )$
\ref{\bala}\ref{\hall}\foot{
Explicitely: If $g\in SU(2)$, it is represented in the
left--regular representation by an operator $U_g$ which acts
on functions on $SU(2)$ itself like: $(U_g F)(g')= F(g^{-1}g')$. If
we choose the Euler angles $\{\phi^a\}$ as coordinates on the
group, the generators of this representation will be some linear
differential operators $\hat T_{Li}=E^a_i(\phi ){\partial\over \partial
\phi^a}$, which give on irreducible representations
 $\hat T_{Li}{\cal D}^j(g)=-T^j_i{\cal D}^j(g)$.
Analytically continuing to $SL(2,\Complexes )$,
$ \phi^a \to \varphi^a+i\eta^a,\ E^a_i\to e_i^a+ih_i^a$; the
continued  generators can be grouped into:
$$ \hat T^c_{Li}=\half (e_i^a+ih_i^a)({\partial\over \partial
\varphi^a}-i{\partial\over \partial\eta^a}) \ ;
\quad{\hat{\bbar T^c}}_{Li}=\half
(e_i^a-ih_i^a)({\partial\over
\partial \varphi^a}+i{\partial\over \partial\eta^a}) $$
which satisfy separately the algebra and commute with each
other. Continuing the irreducible representations to
$SL(2,\Complexes )$, one has
$ \hat T^{c}_{Li}{\cal D}^j(g)=-T^j_i{\cal D}^j(g),
\ \hat T^{c}_{Li}\bbar{\cal D}^j(g)=0$.\nl
 What I am saying is that $\hat S^i=-\hat T^{c}_{Li}$.
 }:
\eqn\xvi{-\hat S^i\Psi (..,g,..)=
\Psi (..,-\tau^ig,..)}
A basis for the wavefunctions is provided by the
"net states"\carlolee :
given  a net in the dual lattice, (the lattice made of  links that
cross our triangles), choose a representation $(j_l,0),\ j_l=0,
\half ,1,\ldots $ of $SL(2,\Complexes )$ for every link $l$ in the
net, and an appropriate invariant tensor $C(\{ j\})$ for every
vertex. Then a gauge invariant, holomorphic wavefunction will be
given by:
\eqn\xvii{\Psi =\sum C(\{ j\})\prod_{l}{\cal D}^{j_l}(g_l)  }
The scheme is however ruined because it cannot have an
acceptable scalar product, i.e. one such that the
$\hat S^i$ are hermitian, and/or $S^2$ is positive.

Suppose one takes as scalar product, starting from the Haar
measure $d\mu_H(g)$ on $SL(2, \Complexes )$:
\eqn\xviii{<\Psi |\Phi >=\int f(g)\;\bbar\Psi (g)\;\Phi
(g)\;d\mu_H(g) }
with some $f(g)$ to be determined. Then since $\int \hat S^i
(..)d\mu_H(g)=0$, and $\hat S^i \bbar\Psi (g)=0$ (see the previous
footnote):
\eqn\xix{\int f(g)\;\bbar\Psi (g)\;\hat S^i\Phi (g)\; d\mu_H(g)=
\int f(g)\;\hat{\bbar S^i}\Psi (g)\;\Phi (g)\; d\mu_H(g)\
\Longrightarrow\ \hat S^i f(g)= \hat{\bbar{S^i}} f(g) }
but this cannot be, because using the commutation relations:
\eqn\xx{\hat S^i f=-\epsilon_{ijk}\hat S^j\hat S^k f =
-\epsilon_{ijk}\hat S^j\hat{\bbar {S^k}} f=-\epsilon_{ijk}\hat{\bbar{S^k}}\hat
S^j f=
-\epsilon_{ijk}\hat{\bbar{S^k}}\;\hat{\bbar{S^j}} f=-\hat{\bbar{S^i}} f}
So the search for a measure fails. A similar argument shows
that it fails also with the more general ansatz for the
scalar product (suggested by Carlo Rovelli)
\eqn\xxi{<\Psi |\Phi >=\int f(g_1,g_2)\;\bar\Psi (g_1)\;\Phi
(g_2)\;d\mu_H(g_1)d\mu_H(g_2) }
where we would require $\hat S^i_2 f=\hat{\bbar {S^i_1}} f$.

On the contrary, the same argument was successful with linearized
gravity, and for the $(q,z=q-ip)$ oscillator \ars\abhay . In this
latter case one sets:
\eqn\xxii{\hat z \psi (z)=z\psi (z)\ ,\ \ \hat q\psi
(z)={\partial\psi\over\partial z}\qquad ;\qquad \hat q^\dagger
=\hat q\ ,\quad \hat z^\dagger = -\hat z +2\hat q }
and finds that these (abelian) reality conditions do have a
solution:
\eqn\xxiii{<\psi |\phi >=c\int d^2z\exp\big(-\quart
(z+z^*)^2\big)\psi (z)^*\phi (z)}
In our case the $i$ factor spoils the whole scheme.

This failure is somewhat disheartening, and would semm to block
further progress in this direction.

\bigskip\noindent{\bf 5. Conclusions.  }\medskip\nobreak\medskip
I have tried to develop a discretized canonical theory of
gravity in the spirit of Regge calculus, and that means keeping as
much as possible of the underlying geometric structure of the
theory. Inevitably many of the original difficulties remain, plus
new ones which may be artefacts of the discretization. The
discussion of \S 4 makes perhaps clear that the
"geometodynamic" alternative is far too complicated to be
practicable. In \S 5 it has been shown that a direct use of the
Ashtekar-like variables does not lead to a sensible quantum
theory. Some suitable middle ground should be found, perhaps
restricting the configuration space with a careful analysis of
the other constraints.

I would like to thank Annalisa Marzuoli, Carlo Rovelli and especially Ruth M.
Williams for discussing the content of this paper with me.

\vskip 0.5truein
\centerline {\bf References}
\nobreak \parindent=0pt \parskip=3pt \singlespace
\medskip  \noindent
\item{\gitwo} G. Immirzi, \CQG{11} (1994) 1971.
\item{\leerent} P. Renteln and L. Smolin, \CQG{6} (1989) 275.
\item{\renate} R. Loll,\NP{B444} (1995) 619 (gr-qc/9502006),
 \PRL{75} (1995) 3048 (gr-qc/9506014), gr-qc/9511080.
\item{\regge} T. Regge, \NC{19} (1951) 558.
\item{\abhay} A. Ashtekar, "Non--perturbative canonical quantum
gravity", notes prepared in collaboration with R. Tate,
 Singapore: World scientific, 1991
\item{\ars} A. Ashtekar, C. Rovelli and L. Smolin, \PR{D44} (1991)
1740.
\item{\carlolee} C. Rovelli and L. Smolin, \NP{B442} (1995) 593, gr-qc/9505006.
\item{\abhayrenate} A. Ashtekar and R. Loll, gr-qc/9405031
\item{\nando} J. F. Barbero, \PR{D51} (1995) 5498, 5507.
\item{\thooft} G. 't Hooft, \CQG{10} (1993) 1653.
\item{\fad} A. Yu. Alekseev and L. D. Faddeev, Commun.Math.Phys
141 (1991) 413.
\item{\frolich} J. Fr\"olich, preprint IHES 1983 reprinted in:
Non perturbative quantum field theory (World Scientific,
Singapore, 1992).
\item{\ruth} T. Piran and R. M. Williams, \PR{D33} (1986) 1622.
\item{\bala} A.P. Balachandran and G.C. Trahern, Lectures on group
theory for physicists, Bibliopolis, Naples 1984.
\item{\hall} B.C. Hall, Journal of Functional Analysis {\bf 122}
(1994) 103.

\end